\documentclass[a4paper]{jpconf}
\usepackage[dvipdfmx]{graphicx}
\usepackage{bm}
\usepackage{epstopdf}
\usepackage{amsmath,amssymb}
\usepackage{float}
\usepackage{wrapfig}
\usepackage{layout}
\usepackage{color}
\usepackage[english]{babel}
\usepackage{braket}
\newcommand{\ii}{\mathrm{i}}
\newcommand{\F}{\mathrm{F}}
\newcommand{\dd}{\mathrm{d}}
\newcommand{\od}{\mathrm{od}}

\newcommand{\imp}{\mathrm{imp}}

\newcommand{\M}{\mathrm{M}}

\newcommand{\n}{\mathrm{n}}

\def\leftdef{\mathrel{\mathop:}=}

\begin{document}
\title{Charged and uncharged vortices in quasiclassical theory}

\author{Yusuke Masaki$^{1}$ and Yusuke Kato$^{1,2}$}

\address{
$^{1}$Department of Physics, The University of Tokyo, Bunkyo, Tokyo 113-0033, Japan
$^{2}$Department of Basic Science, The University of Tokyo, Meguro, Tokyo 153-8902, Japan}

\ead{masaki@vortex.c.u-tokyo.ac.jp}

\begin{abstract}
The charging effect of a superconducting vortex core is very important for transport properties of superconducting vortices. The chiral p-wave superconductor,  known as a topological superconductor (SC), 
has a Majorana fermion in a vortex core and the charging effect has been studied using microscopic Bogoliubov--de Gennes (BdG) theory.  According to calculations based on the BdG theory, one type of the vortex is charged as well as the vortex of the s-wave SC, while the other is uncharged. We reproduce this interesting charging effect using an augmented quasiclassical theory in chiral p-wave SCs, by which we can deal with particle-hole asymmetry in the quasiclassical approximation.
\end{abstract}

\section{Introduction}
Electronic bound states in a superconducting vortex core, which is called the Caroli--de Gennes--Matricon (CdGM) mode~\cite{Caroli1964} or Andreev bound state~\cite{andreev1965thermal, PhysRevB.54.13222}, cause a lot of fascinating phenomena. For example, the disorder-scattering bound states result in the viscous flow of the vortex, while the particle-hole asymmetry of the CdGM mode as well as other origins of the particle-hole asymmetry leads to the Hall effect of the flux flow states~\cite{Larkin1986,Blatter1994, Kopnin200107}. 
The charging effect of the vortex core is another consequence of the particle--hole asymmetry of the CdGM mode. Recently there is another  interest in the zero energy bound state in a vortex core of a topological superconductor, called the Majorana state~\cite{RevModPhys.82.3045, RevModPhys.83.1057}. Majorana bound states obey the non-Abelian statistics and enable a fault tolerant quantum computation~\cite{RevModPhys.80.1083}, which will be done by exchanging the vortices~\cite{IvanovStephanovichZhmudskii1990}. Thus deeper understanding  of the flux flow states of topological superconductors is highly required from both fundamental and applicational viewpoints.

As we mentioned above, the particle-hole asymmetry can be understood as a charging effect of the vortex core~\cite{Khomskii1995, PhysRevLett.77.566}, which has been studied using microscopic Bogoliubov--de Gennes (BdG) theory~\cite{Hayashi1998, PhysRevB.65.014504} and very recently done using augmented quasiclassical theory~\cite{Ueki2016, Ohuchi2017}. The charging effect may be regarded as the main mechanism leading to the Hall effect of the flux flow state~\cite{Khomskii1995}. 
It is numerically prohibitive to study the non-equilibrium phenomena with use of  
the BdG theory or Gor'kov theory. Because of this, the study of the flux-flow state has been done using the time dependent Ginzburg--Landau equation~\cite{Schmid1966}, the Usadel equation~\cite{Usadel1970}, and the Eilenberger equation~\cite{Eilenberger1968a, larkin1969quasiclassical}, the former two of which are valid only for dirty superconductors. The Eilenberger theory is a quasiclassical theory and a very powerful tool to study the flux flow state, but the conventional one is always particle-hole symmetric.
Taking the next leading order of the quasiclassical approximation into account, we can deal with the charging effect and thereby the Hall state in the quasiclassical regime. Kopnin developed the convenient kinetic theory with respect to the Gauge transformation within the Wigner representation~\cite{Kopnin1994}, which was originally given in the normal state~\cite{Levanda1999}.  Subsequently Kita has improved the form of the phase transformation~\cite{PhysRevB.64.054503}.

In this paper we apply this theory to the chiral p-wave SCs~\cite{Mackenzie2003, MaenoKittakaNomuraYonezawaIshida2012}, and study the charging effect.
Here we show that the striking difference between two types of vortices in chiral p-wave SCs can be obtained also in this framework, which is known by  BdG theory~\cite{PhysRevB.65.014504}. 
Our scheme can be extended to systems with disorder in non-equilibrium and thus we believe that our result in the present work becomes a first step toward consistent understanding of the charging effect and Hall effect in chiral p-wave SCs. 

In this paper, we use the unit that $k_{\mathrm{B}} = \hbar =1 $.

\section{Formulation}
Green's function is defined using the Nambu spinor $\vec{\Psi}(\bm{r},\tau) = ^{t}\![\psi_{\uparrow}(\bm{r},\tau),\psi_{\downarrow}^{\dagger}(\bm{r},\tau)]$  as
\begin{eqnarray}
\widehat{G}^{\M} (\bm{r}_{1},\bm{r}_{2},\omega_{n}) 
=
\begin{bmatrix}
G^{\M}(\bm{r}_{1},\bm{r}_{2},\omega_{n})  & F^{\M}(\bm{r}_{1},\bm{r}_{2},\omega_{n})\\
-\bar{F}^{\M}(\bm{r}_{1},\bm{r}_{2},\omega_{n})  & \bar{G}^{\M}(\bm{r}_{1},\bm{r}_{2},\omega_{n}) 
\end{bmatrix}=\int_{0}^{\beta} \dd\tau e^{\ii \omega_{n} \tau} \widehat{\tau}_{3}
\begin{bmatrix}
\braket{ \vec{\Psi}(\bm{r}_{1},\tau)\vec{\Psi}^{\dagger}(\bm{r}_{2})}
\end{bmatrix}.
\end{eqnarray}
Here $\omega_{n}$ and $T$ are the fermion Matsubara frequency and temperature, respectively, and $\beta=1/T$ is  the inverse temperature. 
The symbol $\widehat{\cdot}$ denotes the 2 by 2 matrix, and $\widehat{\tau}_{3}$ is the third component of the 2 by 2 Pauli matrix.
The Fourier transform with respect to the relative spatial and temporal coordinates leads to the Wigner representation.  In this paper, we only consider the system in equilibrium, 
which means that Green's function depends only on the relative imaginary time. When we discuss the gauge transformation it is convenient to do the phase transformation introduced in Refs.~\cite{Kopnin1994, PhysRevB.64.054503}. In this paper, we consider the phase transformation only in the space and vector potential, and consider the partially fixed gauge, where the electric field is described by the scalar potential only. The same transformation was done in Refs.~\cite{Ueki2016, Ohuchi2017}.
Within this scheme, we restrict ourselves to the equilibrium situation, namely, there are no time dependence even in the gauge field.

The Green's function in the space-momentum mixed representation and thereby the quasiclassical Green's function are introduced by the Fourier transform with the relative coordinate $\bm{r} = \bm{r}_{1}-\bm{r}_{2}$ and the energy integration: 
\begin{eqnarray}
\widehat{g}^{\M}(\bm{k}_{\F},\bm{R};\ii\omega_{n})
&=& \oint_{C_{\mathrm{qc}}} \dfrac{\dd \xi_{\bm{k}} }{\ii \pi}
\widehat{G}^{\M}(\bm{k},\bm{R};\ii \omega_{n})\nonumber \\
&=& \oint_{C_{\mathrm{qc}}} \dfrac{\dd \xi_{\bm{k}} }{\ii \pi}\int \dd \bm{r}
e^{\ii I(\bm{R},\bm{r}_{1})\widehat{\tau}_{3}}
\widehat{G}^{\M}(\bm{r}_{1},\bm{r}_{2};\ii \omega_{n})
e^{-\ii I(\bm{R},\bm{r}_{2})\widehat{\tau}_{3}}
e^{-\ii \bm{k}\cdot\bm{r}},
\end{eqnarray}
where $C_{\mathrm{qc}}$ is the contour used in Ref.~\cite{Eilenberger1968a} and the phase $I(\bm{R},\bm{r}_{1})$ is defined as
$I(\bm{R},\bm{r}_{1}) = -\frac{e}{c}\int_{\bm{r}_{1}}^{\bm{R}}  \dd \bm{s} \cdot \bm{A}(\bm{s})$ with vector potential $\bm{A}$, the center-of-mass coordinate $\bm{R} = (\bm{r}_{1} + \bm{r}_{2})/2$, electron charge $e$ and the speed of light $c$. We describe the components of the quasiclassical Green's function as $\widehat{g}^{\M} = \begin{bmatrix}g & f\\ -\bar{f} & \bar{g}\end{bmatrix}$.

We obtain the Eilenberger equation by the following steps: (i) Expand the left- and right-Gor'kov equations for the above phase transformed Green's function up to the second order in the gradient expansion (ii) Subtract the right equation from the left equation (iii) Integrate the obtained equation along the above contour with the approximation that the momentum dependence of the self energy and pair potential terms can be treated by considering only the Fermi momentum $\bm{k}_{\F}$. 
The augmented Eilenberger equation is finally obtained  as follows:
\begin{equation}
[\ii\omega_{n}\widehat{\tau}_{3} + \widehat{\sigma}^{\M},\widehat{g}^{\M}]_{\star}
+\ii \bm{v}_{\bm{k}_{\F}}\cdot \partial_{\bm{R}}\widehat{g}^{\M} 
+
\dfrac{e}{2} \left(\bm{v}_{\bm{k}_{\F}} \times \bm{B}\right)
\left(\ii \dfrac{\partial}{\partial \bm{k}_{\F,\parallel}}\right)
\left\{\widehat{\tau}_{3}, \widehat{g}^{\M}\right\} = 0 \label{eq-Kita}
\end{equation}
with the self energy consisting of the pair potential $\Delta$ and the impurity self-energy $\widehat{\sigma}_{\imp}^{\M}$
\begin{equation}
\widehat{\sigma}^{\M} = 
\begin{bmatrix}
\sigma_{\dd} & \sigma_{\od} \\
\bar{\sigma}_{\od} & \bar{\sigma}_{\dd}
\end{bmatrix}
=
\begin{bmatrix}
0 & \Delta(\bm{k}_{\F},\bm{R}) \\
-\Delta^{*}(\bm{k}_{\F},\bm{R})& 0
\end{bmatrix} + \widehat{\sigma}_{\imp}^{\M}
\end{equation}
and
\begin{equation}
\partial_{\bm{R}} = 
\begin{cases}
\bm{\nabla} &\text{on}~g,\bar{g}, \sigma_{\dd},\bar{\sigma}_{\dd}\\\vspace{0.5em}
\bm{\nabla} -\dfrac{2\ii e}{c}\bm{A}(\bm{R}) &\text{on}~f,\sigma_{\od}\\
\bm{\nabla} +\dfrac{2\ii e}{c}\bm{A}(\bm{R}) &\text{on}~\bar{f}, \bar{\sigma}_{\od}.
\end{cases}
\end{equation}
The braces in the last term of Eq.~\eqref{eq-Kita} denote the anticommutator.
The subscript $\parallel$ in the momentum derivative stands for the tangential component of the momentum to the Fermi surface.
The subscript $\star$ of the first term in Eq.~\eqref{eq-Kita} means that the product of the commutation relation is replaced by the Moyal product in the spatial part ($\bm{r}-\bm{k}$ derivatives), whose definition is given by
\begin{eqnarray}
[A,B]_{\star}
\!\!\!&\equiv&\!\!\!
A\star B
-B\star A, \\
A\star B 
\!\!\!&\equiv&\!\!\!
\lim_{\bm{R}^{\prime} \to \bm{R}}\lim_{\bm{k}_{\F}^{\prime}\to \bm{k}_{\F}
}\exp\left[\dfrac{\ii}{2}\left(\dfrac{\partial}{\partial \bm{k}_{\F,\parallel}^{\prime}}\cdot \partial_{\bm{R}}
-\dfrac{\partial}{\partial \bm{k}_{\F,\parallel}}\cdot \partial_{\bm{R}^{\prime}}
\right)\right]
A(\bm{k}_{\F},\bm{R};\ii\omega_{n}) B(\bm{k}_{\F}^{\prime},\bm{R}^{\prime};\ii\omega_{n}).
\end{eqnarray}
In the above expression, we retain the terms up to the first order.
We consider the case where the Fermi velocity $\bm{v}_{\F}$ is given by $\bm{k}_{\F}/m$.

We write down the augmented quasiclassical equation order by order with respect to the quasiclassical parameter. 
In the following, we neglect the magnetic field for simplicity. 
The Green's function as well as the self energy and the pair potential terms can be expanded as 
$\mathcal{O} = \mathcal{O}_{0} + \mathcal{O}_{1} + \cdots$, where $\mathcal{O}$ denotes any of $\widehat{g}^{\M}$, $\widehat{\sigma}^{\M}$, and $\Delta$. 
The lowest order equation is reduced to the conventional Eilenberger equation, which can be solved using the Riccati transformation supplemented with the gap equation and the explicit form of the self energy.
\begin{eqnarray}
2\ii \omega_{n} f_{0} + \ii \bm{v}_{\F} \cdot \partial_{\bm{R}} f_{0} - 2\Delta_{0} g_{0}
+2\ii \Gamma_{\n}[f_{0}\braket{g_{0}}_{\F}-\braket{f_{0}}_{\F}g_{0}] = 0 ,\\
g_{0} = \mathrm{sgn}(\omega_{n} )\left[1- f_{0}\bar{f}_{0}\right]^{1/2},\\
\Delta_{0}(\bm{k}_{\F},\bm{R}) = g \nu_{0}(\xi_{\F})\ii \pi T \sum_{n}\Braket{V_{\bm{k}_{\F},\bm{k}_{\F}^{\prime}}^{L_{z}}f_{0}^{\M}(\bm{k}_{\F}^{\prime},\bm{R};\ii\omega_{n})}_{\F^{\prime}}, \\
\widehat{\sigma}_{\imp,0}^{\M}(\bm{k}_{\F},\bm{R};\ii\omega_{n}) = \ii\Gamma_{\n}\braket{\widehat{g}_{0}^{\M}(\bm{k}_{\F}^{\prime},\bm{R};\ii\omega_{n})}_{\F^{\prime}},
\end{eqnarray}
where $V_{\bm{k}_{\F},\bm{k}_{\F}^{\prime}}^{L_{z}} = 1$ for s-wave SC ($|L_{z}|= 1$) and  $2\cos (\alpha-\alpha^{\prime})$ for chiral p-wave SCs ($|L_{z}| = 2 $ or 0). Note that $L_{z}$ describes the total angular momentum of the Cooper pair~\cite{KatoHayashi2001, Masaki2015, Masaki2016}. We have introduced $\alpha$ as a direction of the Fermi momentum in two dimensional (2D) space.
The second equation describes the normalization condition for the zeroth order quasiclassical Green's function. We note that $\braket{o(\bm{k}_{\F})}_{\F}$ denotes the average over the 2D Fermi surface defined by $(2\pi)^{-1}\int_{0}^{2\pi} \dd \alpha  o(\bm{k}_{\F})$, and $\nu_{0}(\xi_{\F})$ and $g$ are the 2D density of states per spin at the Fermi level $\xi_{\F}$ and the coupling constant, respectively. The coupling constant is related to the transition temperature $T_{\mathrm{c}}$ as 
\begin{equation}
\dfrac{1}{g \nu_{0}(\xi_{\F})} = \ln \dfrac{T}{T_{\mathrm{c}}} + 2\pi \sum_{0\le n\le N_{\mathrm{c}} }\dfrac{1}{2n+1}
\end{equation}
with a cut-off number $N_{\mathrm{c}}$. The impurity scattering rate in the normal state is described by $\Gamma_{\n}$.

The first order equation consists of two closed parts: One is an equation for $g_{1} +\bar{g}_{1}$, and the other is a coupled equation for $g_{1} - \bar{g}_{1}$, $f_{1}$, and $\bar{f}_{1}$.
The charging effect, as we see next, is described by $g_{1} + \bar{g}_{1}$, which obeys
\begin{eqnarray}
\dfrac{1}{2}\ii v_{\F} \dfrac{\partial (g_{1} + \bar{g}_{1})}{\partial s} 
=
\hspace{-1.5em}&&
\dfrac{\ii \hat{e}_{b}}{2k_{\F}}\cdot\left[2\left(\partial_{\bm{R}}g_{0} \dfrac{\partial \sigma_{\dd,0}}{\partial \alpha}-\partial_{\bm{R}}\sigma_{\dd,0} \dfrac{\partial g_{0}}{\partial \alpha}
\right)\right.
\nonumber \\
&&\hspace{3.5em}+\left.\partial_{\bm{R}}\sigma_{\od,0} \dfrac{\partial \bar{f}_{0}}{\partial \alpha}
-\partial_{\bm{R}}\bar{\sigma}_{\od,0} \dfrac{\partial f_{0}}{\partial \alpha}
-\dfrac{\partial \sigma_{\od,0}}{\partial \alpha} \partial_{\bm{R}}\bar{f}_{0}
+\dfrac{\partial \bar{\sigma}_{\od,0} }{\partial \alpha}\partial_{\bm{R}}f_{0}
\right].
\end{eqnarray}
Here $\hat{e}_{b} \leftdef \hat{e}_{z} \times \hat{e}_{s}$ with $\hat{e}_{s} = \bm{v}_{\F} /v_{\F}$ and $v_{\F} =|\bm{v}_{\F}|$, and $s$ is the cooridinate along $\hat{e}_{s}$.

Next we see how the charging effect is obtained. The 2D charge density is given by~\cite{Eliashberg1972, Serene1983}
\begin{equation}
e(n(\bm{R}) - n_{\n})
=- e\ii \pi T \sum_{n}  \nu_{0}(\xi_{\F})
\Tr\left[\Braket{\widehat{g}^{\M}}_{\F}-\Braket{\widehat{g}_{\n}^{\M}}_{\F}\right]
 - 2\nu_{0}(\xi_{\F})e^{2}\Phi(\bm{R})
,\label{eq-charge}
\end{equation}
where the scalar potential $\Phi$ on the right-hand side comes from the high-energy contribution.
Note that if we do the phase transformation involved with the scalar potential, particularly in the Keldysh formalism, we do not have the scalar potential explicitly in the charge density, instead we have the electric field term in the Eilenberger equation~\cite{Kopnin200107, ArahataKato2014}.
The left-hand side is described as the difference in the charge density between the superconducting state in question and the normal state without any gauge fields.

The charge expression, of course, includes the scalar potential, and hence we have to solve the complementary Poisson equation for the thickness $d$, given by
\begin{align}
\nabla^2 \Phi(\bm{R}) = -\dfrac{4\pi e}{d} (n(\bm{R}) - n_{\n}), \label{eq-poisson}
\end{align}
or
\begin{equation}
\left[\nabla^{2} - \lambda_{\mathrm{TF}}^{-2}\right]\Phi(\bm{R}) = -  \lambda_{\mathrm{TF}}^{-2}\dfrac{2\pi T}{e} \sum_{\omega_{n} > 0 }\Braket{\mathrm{Im} \left[\dfrac{g_{1}+\bar{g}_{1}}{2}\right](\bm{k}_{\F},\bm{R}; \ii \omega_{n})}_{\F},
\end{equation}
where we use Eq.~\eqref{eq-charge} and the definition of the Thomas--Fermi length
$\lambda_{\mathrm{TF}} = [d/(8\pi\nu_{0}(\xi_{\F})e^{2})]^{1/2}$.
The second term in the left-hand side describes the screening effect. First we solve this Poisson equation through the perturbative Green's function, then we obtain the charge density from Eq.~\eqref{eq-charge} or~\eqref{eq-poisson}. 

\section{Numerical results}
In this section we show the numerical results for two types of vortices of chiral p-wave SCs.
Two types of vortices are characterized by the transformation properties; the pair potential for one type of the vortex acquires the additional phase $2\theta$ under  the simultaneous  $\theta$ rotation of real and momentum spaces, and we label it as $L_{z} = 2$. The pair potential for the other type of the vortex is invariant under the same transformation, and we label it as $L_{z}= 0$. The charge density is normalized by $2|e|T_{\mathrm{c}}\nu_{0}(\xi_{\F})$ and use the notation $\delta \bar{\rho}(R) = e(n(R)-n_{\n})/(2|e|T_{\mathrm{c}}\nu_{0}(\xi_{\F}))$. 
Also we define the integral of the charge density from the origin to radius $R$ as
\begin{equation}
\bar{Q}(R)\xi_{0}^{2} = \int_{0}^{R} \dd R^{\prime} R^{\prime} \delta \bar{\rho}(R^{\prime}), \label{eq-total}
\end{equation}  
where $\xi_{0} = v_{\F}/\Delta_{\infty}$ and $\Delta_{\infty}=\lim_{R\to\infty}|\Delta(\bm{k}_{\F},\bm{R})|$ are the coherence length and  the magnitude of the gap in the bulk region, respectively.
$\bar{Q}(R)$ can be interpreted as the actual charge within the radius $R$ and we have confirmed that the neutrality of the whole system holds ($\bar{Q}(R\to \infty)=0$), 
although we do not show any data about this in the following.
We set $T$ to $0.2T_{\mathrm{c}}$, 
the quasiclassical parameter $(k_{\F}\xi_{0})^{-1}$ to $0.01$,  and $\Gamma_{\n}$ to 0,
in all results presented below.

\begin{figure}[t]
\includegraphics[width = 40em]{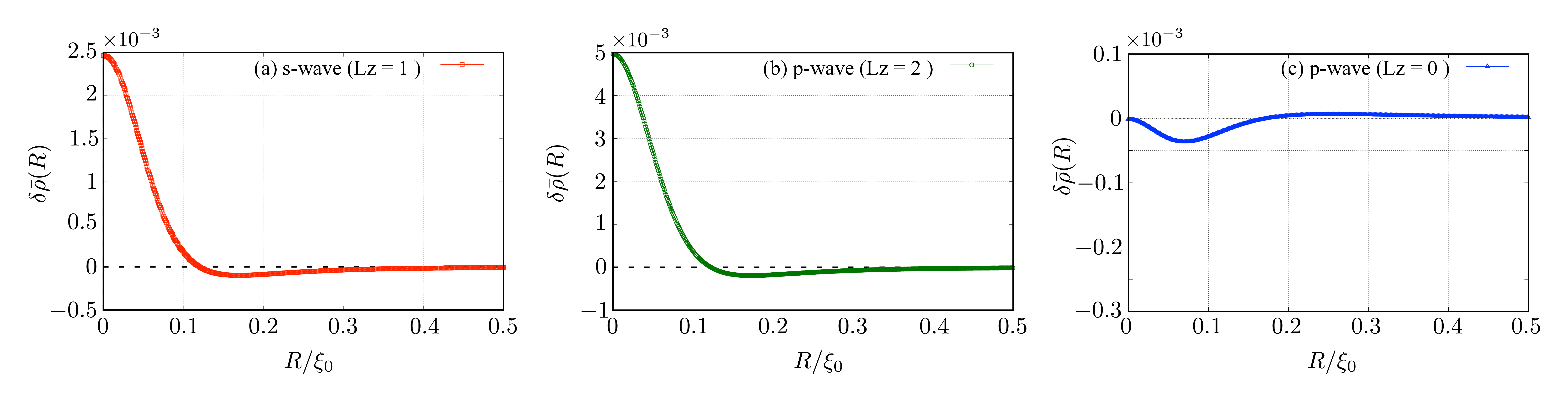}
\caption{Charge density profiles for (a) the s-wave SC, (b) the chiral p-wave SC with $L_{z}=2$, and (c) the chiral p-wave SC with $L_{z}=0$.  }
\label{fig-charge}
\end{figure}

We show the charge profile and the total charge as functions of the radial coordinate. We set the Thomas--Fermi length $\lambda_{\mathrm{TF}}$ to $0.01\xi_{0}$. 
First we neglect the effects of the induced components in chiral p-wave SCs, which are discussed later. 
Figures~\ref{fig-charge}(b) and \ref{fig-charge}(c) are for $L_{z}=2$ and $L_{z}=0$, respectively.
Note that the s-wave SC is the case of $L_{z} = 1$, which we show in Fig.~\ref{fig-charge}(a) as a reference. We see that the behavior of $L_{z} =2$ is similar to that of $L_{z}=1$, 
while there is conspicuous suppression in the case of $L_{z} = 0$. 
This can be understood as follows. 
The spatial derivative in the Moyal product includes radial and angular directions and the latter is proportional to the Cooper pair angular momentum $L_{z}$. 
When $L_{z}\neq 0$, the contribution from the angular component is dominant over that from the radial component.
In the case of the zero angular momentum state $L_{z}=0$, this contribution vanishes, and this is the reason why the vortex charge of $L_{z}=0$ is extremely small compared with those of  $L_{z}=1$ and $2$.
In addition, the charge of $L_{z}=2$ is almost twice as large as that of $L_{z}=1$. 
When we do not take account of the induced component, the gap functions as well as the Green's functions are quite similar except the difference of phase factors related to the angular momentum. 

\begin{figure}[t]
\centering
\includegraphics[width = 30em]{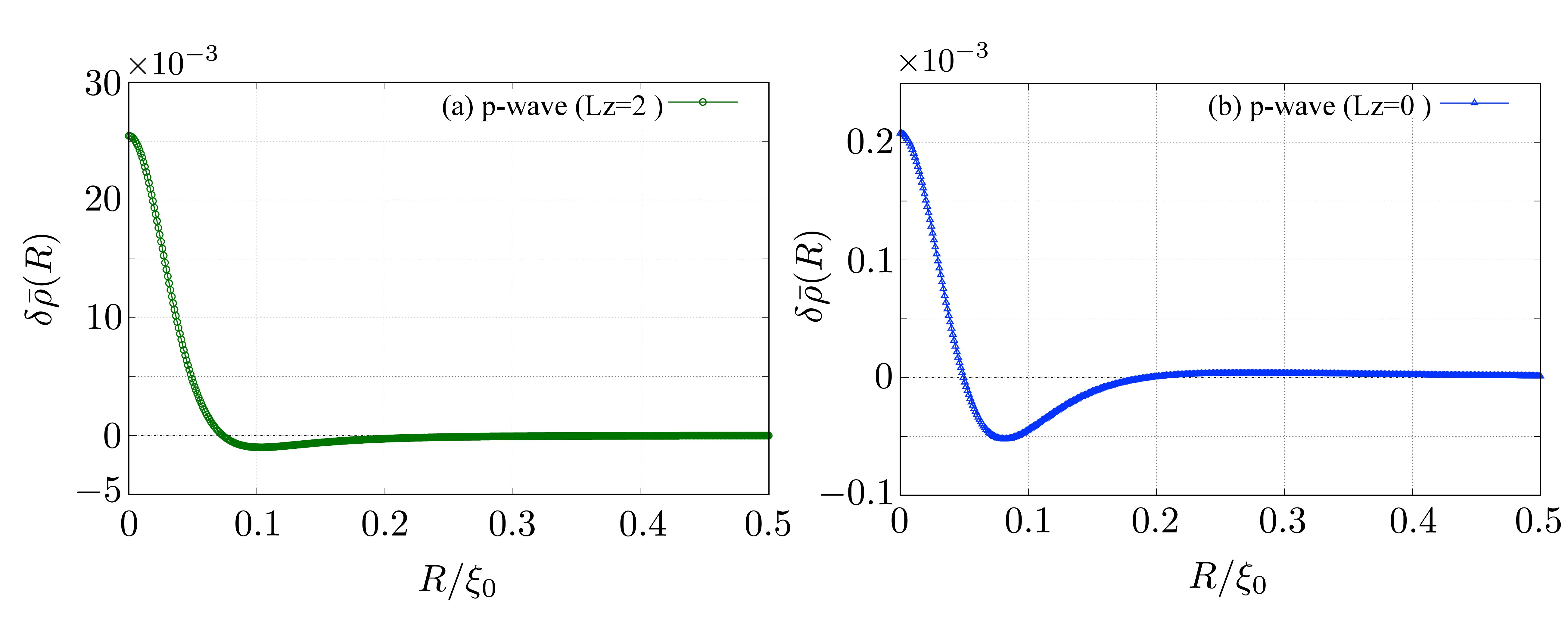}
\caption{
The charging profiles of chiral p-wave SCs with consideration of induced components of gap function. Note that the ranges of the vertical axes are different from those in Fig.~\ref{fig-charge}. 
}
\label{fig-induced}
\end{figure}
Next we discuss the effects of induced components. The profiles of the charge densities are shown in Figs.~\ref{fig-induced}(a) and (b). 
Induced components do not affect the qualitative properties, namely, 
the vortex of $L_{z}=2$ is charged while the vortex of $L_{z}=0$ is hardly charged. 
There is, however, an interesting and quantitative difference in the case of $L_{z}=2$. Figure~\ref{fig-induced} (a) shows that the charge density at the origin becomes five times  as large as that of Fig.~\ref{fig-charge} (b) and thus the vortex charge of $L_{z}=2$ is an order of magnitude greater than that of  the s-wave SC. The dominant contribution is still from the angular derivative, but the changes of the gap function and Green's function cause this enhancement.

\section{Conclusion} 
We have studied the vortex charging effect for chiral p-wave SCs using the augmented quasiclassical theory. There are two types of vortices in the chiral p-wave SCs, and in one case where $L_{z} = 0$,  the charging effect is noticeably suppressed. 
Consistently to the calculation using the BdG theory, the dominant contribution is from the angular component of the spatial derivative, which vanishes for $L_{z}=0$. 
The total angular momentum $L_{z}$ describes the cancellation of two phases in momentum and real spaces.
The induced components affect the gap functions and the quasiclassical Green's functions, and thereby lead to the enhancement of the vortex charge in the case of $L_{z}=2$. 
The effect of impurities should be discussed in subsequent work.

\ack
We would like to thank Y.~Tsutsumi for helpful discussions. 
This work was supported by a Grant-in-Aid for JSPS Fellows (Grant No.~16J03224) and JSPS KAKENHI Grant Number 15K05160. 
Y. M. was supported by the Program for Leading Graduate Schools, the Ministry of Education, Culture, Sports, Science and Technology, Japan.
\section*{References}
\bibliographystyle{iopart-num}
\bibliography{library}

\end{document}